\documentstyle[12pt,epsf]{article}
\def\journal#1, #2, #3, #4 { {\sl #1~}{\bf #2~}(#3) #4 }

\def\cmp{\journal Comm. Math. Phys., }

\def\np{\journal Nucl. Phys., }

\def\pl{\journal Phys. Lett., }

\catcode`\@=11
\def\marginnote#1{}
\newcount\hour
\newcount\minute
\newtoks\amorpm
\hour=\time\divide\hour by60
\minute=\time{\multiply\hour by60 \global\advance\minute
by-\hour}\edef\standardtime{{\ifnum\hour<12
\global\amorpm={am}%
        \else\global\amorpm={pm}\advance\hour by-12 \fi
        \ifnum\hour=0 \hour=12 \fi
        \number\hour:\ifnum\minute<10
0\fi\number\minute\the\amorpm}}
\edef\militarytime{\number\hour:\ifnum\minute<10
0\fi\number\minute}

\def\draftlabel#1{{\@bsphack\if@filesw {\let\thepage\relax
   \xdef\@gtempa{\write\@auxout{\string
      \newlabel{#1}{{\@currentlabel}{\thepage}}}}}\@gtempa
   \if@nobreak \ifvmode\nobreak\fi\fi\fi\@esphack}
        \gdef\@eqnlabel{#1}}
\def\@eqnlabel{}
\def\@vacuum{}
\def\draftmarginnote#1{\marginpar{\raggedright\scriptsize\tt#1}}
\def\draft{\oddsidemargin -.5truein
        \def\@oddfoot{\sl preliminary draft \hfil
        \rm\thepage\hfil\sl\today\quad\militarytime}
        \let\@evenfoot\@oddfoot \overfullrule 3pt
        \let\label=\draftlabel
        \let\marginnote=\draftmarginnote

\def\@eqnnum{(\theequation)\rlap{\kern\marginparsep\tt\@eqnlabel}%
\global\let\@eqnlabel\@vacuum}  }

\def\numberbysection{\@addtoreset{equation}{section}
        \def\theequation{\thesection.\arabic{equation}}}

\def\underline#1{\relax\ifmmode\@@underline#1\else
 $\@@underline{\hbox{#1}}$\relax\fi}

\catcode`@=12
\relax

\numberbysection
\def\fin{\end{document}}
\def\pish{{\pi\over h}}
\def\hspi{{h\over \pi}}
\def\pishp{{\pi\over h'}}
\def\hspip{{h'\over \pi}}
\def\beq{\begin{equation}}
\def\eeq{\end{equation}}
\def\beqa{\begin{eqnarray}}
\def\eeqa{\end{eqnarray}}
 \def\nnn{\nonumber \\}
\def\sqr#1#2{{\vcenter{\vbox{\hrule height.#2pt
\hbox{\vrule width.#2pt height#1pt \kern#1pt
\vrule width.#2pt}
\hrule height.#2pt}}}}

\def\Jne#1 {J_{#1}^e\, \!}
\def\Jneb#1 {{\overline J}_{#1}^e\, \!}
\def\Jnep#1 {J_{#1}'\, \!^e\, \!}
\def\Jnebp#1 {{\overline J}_{#1}'\, \!  \!^e\, \!}

\def\hhat{{\widehat h}}

\def\Jhat{{\widehat J}}

\def\mhat{{\widehat m}}

\def\chib{{\overline \chi}}

\def\Rhat{{\widehat R}}

\def\nuhat{{\widehat \nu}}

\def\qhat{{\widehat q}}

\def\varpihat{{\widehat \varpi}}

\def\varpihatb{{\widehat {\overline \varpi}}}

\def\varpib{{\overline \varpi}}

\def\mhat{{\widehat m}}

\def\square{\mathchoice\sqr34\sqr34\sqr{2.1}3\sqr{1.5}3}
\def\qed{ $ \quad\square $ }

\def\Jb{{\overline J}}

\def\mb{{\overline  m}}
\def\mub{{\bar \mu}}
\def\zb{{\bar z}}

\def\mhatb{\widehat {\overline S}}

\def\Vt{{\widetilde V}}
\def\Vtb{{\widetilde {\overline V}}}

\def\Jhatb{\widehat {\overline J}}
\def\mhatb{\widehat {\overline m}}

\def\Jgen#1 {  {\underline J_{#1}} }
\def\Jgenp#1 #2 {(J_{#1}+{#2},\Jhat_{#1})}
\def\Jgenm#1 #2 {(J_{#1}-{#2},\Jhat_{#1})}
\def\Jg#1 {J_{#1},\Jhat_{#1}}
\def\Jgp#1 #2 {J_{#1}+{#2},\Jhat_{#1}}
\def\Mgen#1 {{\underline M_{#1}}}

\def\produit#1,#2,#3,#4 {P\Bigl (
[{#1},{#2}]\otimes\{{#3}\},{#4}\Bigr )}
\def\produitscript#1,#2,#3,#4 {P\Bigl (
[{\scriptstyle{#1},{#2}}]\otimes\{{\scriptstyle{#3}}\},{#4}\Bigr
)}
\def\pprod#1,#2,#3,#4,#5 {P\Bigl (
[{#1},{#2}]\otimes[{#3},{#4}],{#5}\Bigr )}
\def\pprodscript#1,#2,#3,#4,#5 {P\Bigl (
[{\scriptstyle{#1},{#2}}]\otimes[{\scriptstyle{#3},{#4}}],{#5}\Bigr
)}

\def\fusV#1,#2,#3,#4,#5,#6 {f_V(
\Jgen{#1} ,
\Jgen{#2} ,
\Jgen{#3} ,
\Jgen{#4} ,
\Jgen{#5} ,
\Jgen{#6} )}

\def\brdV#1,#2,#3,#4,#5,#6 {b_V(
\Jgen{#1} ,
\Jgen{#2} ,
\Jgen{#3} ,
\Jgen{#4} ,
\Jgen{#5} ,
\Jgen{#6} )}

\def\fusxi#1,#2,#3 {f_\xi (\Jgen{#1} ,
\Mgen{#1} ,
\Jgen{#2} ,
\Mgen{#2} ,
\Jgen{#3} )}

\def\ghat{{\widehat g}}

\def\gaghat{{\hat {\bigl \{}}}
\def\gadhat{{\hat {\bigr \}}}}

\def\sixjxi#1,#2,#3,#4,#5,#6 {{\left\{\left . \!\! \,^{#1}_{#2}
\,^{#3}_{#4} \right | \!\, ^{#5}_{#6}\right\}}}
\def\sixje#1,#2,#3,#4,#5,#6 {{\left\{\left\{\left . \!\! \,
^{#1}_{#2}
\, ^{#3}_{#4} \right | \!\, ^{#5}_{#6}\right\}\right\}}}
\def\sixjxihat#1,#2,#3,#4,#5,#6 {{{\gaghat\left . \!\! \,
^{#1}_{#2}
\, ^{#3}_{#4} \right | \!\, ^{#5}_{#6}\gadhat}}}
\def\sixjehat#1,#2,#3,#4,#5,#6 {{\gaghat\gaghat\left . \!\! \,
^{#1}_{#2}
\, ^{#3}_{#4} \right | \!\, ^{#5}_{#6}\gadhat\gadhat}}

\def\epsilonhat{{\widehat \epsilon}}

\def\Jhatb{\widehat {\overline J}}
\def\mhatb{\widehat {\overline m}}
\def\vertex#1,#2,{{{\cal V}^{{#1,#2}}}}
\def\vertexc#1,#2,{{{\cal V}^{{#1,#2}}_{conj}}}
\def\lf#1,#2,{{L_{#1,#2}}}
\def\lfc#1,#2,{{{\bar L}_{#1,#2}}}
\def\ns{{\nu}}
\def\nsb{{\bar \nu}}
\def\cn{a_-}
\def\cb{a_+}

\def\nue{\nu^e{}}
 
\def\chemin#1,#2 {{\left\Uparrow {{#2}\atop {#1}} \right\Uparrow}}
\def\cheminp#1,#2 {{\left\Uparrow {{#2}\atop {#1}} \right\Uparrow}'}
\def\hspi{{h\over \pi}}
\def\pish{{\pi\over h}}
\def\demi{{1\over 2}}

\def\vertexc#1,#2,{{{\cal V}^{{#1,#2}}_{\hbox{\scriptsize conj}}}}
\def\lfc#1,#2,{{L^{\hbox{\scriptsize conj}}_{#1,#2}}}

\def\vertexw#1,#2,{{{\cal W}^{{#1,#2}}}}
\def\vertexcw#1,#2,{{{\cal W}^{{#1,#2}}_{\hbox{\scriptsize conj}}}}
\def\pw {Q}
\def\lfw#1,#2,{{M_{#1,#2}}}
\def\lfcw#1,#2,{{M^{\hbox{\scriptsize conj}}_{#1,#2}}}
\def\nshat {\widehat \nu}

\begin{document}
\tolerance 2000
\hbadness 2000
\begin{titlepage}

\begin{flushright}

LPTENS--95/25, \\
hep-th/9605105, \\
May   1996.
\end{flushright}

\vglue 3   true cm
\begin{center}
{\large \bf
SOLVING THE STRONGLY COUPLED 2D GRAVITY \\ \bigskip
III: STRING SUSCEPTIBILITY AND\\ \bigskip
TOPOLOGICAL N-POINT  FUNCTIONS}
\vglue 1.5 true cm
{\bf Jean-Loup~GERVAIS}\\
\medskip
{\bf Jean-Fran\c cois ROUSSEL}\\
\medskip
{\footnotesize Laboratoire de Physique Th\'eorique de
l'\'Ecole Normale Sup\'erieure\footnote{Unit\'e Propre du
Centre National de la Recherche Scientifique,
associ\'ee \`a l'\'Ecole Normale Sup\'erieure et \`a
l'Universit\'e
de Paris-Sud.},\\
24 rue Lhomond, 75231 Paris CEDEX 05, ~France}.
\end{center}
\vfill
\begin{abstract}
\baselineskip .4 true cm
\noindent
{\footnotesize
We spell out the  derivation of   novel features, put forward  
earlier in a letter, 
 of  two dimensional gravity
in the strong coupling regime, at $C_L=7$, $13$, $19$. Within the 
operator approach previously developed,
 they neatly follow 
from   the appearence of a new  
cosmological term/marginal operator, different from the standard weak-coupling one, 
 that determines the 
world sheet interaction. The corresponding  string susceptibility 
is obtained and found real contrary to the continuation of the KPZ 
formula.  Strongly coupled (topological like) models---only involving zero-mode 
degrees of freedom---are 
solved up to sixth order, using the ward identities which follow from the 
dependence upon the new cosmological constant. They are technically similar to the weakly 
coupled ones, which reproduce the matrix model results, but gravity and matter quantum 
numbers are entangled differently.  

 }
\end{abstract}
\vfill
\end{titlepage}

\section{Introduction}
The truncation of the chiral operator  algebra at $C_L=7$, $13$, $19$ 
 seems to be the key 
to the strong coupling regime of two dimensional gravity\footnote{Evidence 
that a similar mechanism is at work for $W_3$  gravity are 
given in ref.\cite{GR}.}. After 
the first hints\cite{GN5}, 
 the trucations theorems 
were proven  at the level of primaries (for half integer spins) 
in ref.\cite{G3}, the final complete proof being given in a recent paper 
of ours\cite{GR1}. In ref.\cite{GR2}, and  in the present article, we analyse further the physics
of these theories and begin solving the topological 
models proposed in ref.\cite{GR1}, where gravity is in 
the strong coupling regime. 

In the strong coupling regime, the screening charges are
complex, so that the dimensions of  primaries are  not real 
in general. These truncation theorems express the fact that, 
nevertheless,  
at the special values of $C_L$  the operator algebra of 
the chiral intertwinors---that appear as chiral components of the
powers of the Liouville exponentials---has a consistent restriction to 
Verma modules with 
real Virasoro highest  weights. Thus one is led to consider that
strongly coupled two dimensional gravity should be expressible entirely 
in terms  of the corresponding chiral components. This brings in  drastic 
changes with respect to the weakly coupled regime. It rules out 
the Liouville exponentials, which are the well known local fields of the
present description of 2D gravity in the conformal gauge, 
 both classically and at the weakly 
coupled quantum
level\cite{G3}. However, a new set of local fields may be constructed,
which is  consistent with the truncation theorems.   
Then the corresponding physics follows quite naturally, using methods 
very similar to the weak coupling ones. 

In ref.\cite{GR1} we emphasized that the transition from weak to strong 
coupling may be characterized by a deconfinement of chirality. Indeed, 
in the weak coupling regime, the Liouville exponentials may be consistently 
restricted to the sector where left and right chiral components have the 
same Virasoro weight, so that we may say that chirality (of gravity) is confined. 
In this regime, the quantum numbers associated with each screening charge 
may take independent values. In the strong coupling regime, on the contrary, 
the reality condition forces us to link the quantum numbers associated 
with the two screening charges, for each chirality. This is however 
incompatible with the equality between left and right weights, so 
that the (gravity) chirality fluctuates: it is deconfined. 
The main motivation to 
study the strongly coupled regime is of course the building of  the (so called) 
non critical strings. Indeed, the balance of central charges prevents us from 
doing so in the  weak coupling regime. The  actual construction  
is still too complicated beyond the three point functions, but we have set up\cite{GR1,GR2}  
models---which we call topological since their only degrees of freedom are 
zero modes---where gravity is strongly coupled. We shall present details about their solution 
which were left out from ref.\cite{GR2}.  

The article is organised as follows. In section 2 we derive the string susceptibility, 
drawing a parallel with the weak coupling calculation. In section 3, we study the  
three point functions of our topological models. It was derived earlier\cite{GR1} 
only with the same screening choice for the three legs, but we need to treat the 
mixed case. Section 4 is devoted to the derivation of the N point function 
from the Ward identities generated by taking derivatives with respect to 
the strong-coupling cosmological constant. Using Mathematica, we derive  
irreducible  irreducible vertices, up to six legs, which give  
a consistent perturbation theory up to  sixth  order.  This is similar to  earlier studies 
in the weak coupling regime\cite{DFK}  
but we let the screening number fluctuate, contrary to 
this reference. Consistency of the 
perturbative expansion of the topological models 
leads  us to counting rules for its relation with the Feynman 
path integrals, which could not 
be guessed a priori. However,  at the end we verify that the same rules also hold in the 
weak coupling regime, if  the screening numbers are allowed to fluctate. As a matter of 
fact, the two types of topological models (i.e. the continuous version of matrix models, and our 
new ones) are found to be very similar technically, although 
 gravity and matter quantum numbers are entangled completely differently.  

\section{The string susceptibility}
\label{string susceptibility}
\subsection{The weak coupling case revisited}
To begin with, we review the derivation in the weak coupling regime, 
following essentially  
ref.\cite{G5}. We shall  provide details  left out 
previously which  will shed light on the strong coupling case. 
In particular, it 
will be useful to draw a close parallel  with the 
DDK argument. With our notations, the latter follows from the              
following\footnote{
The upper index ``(w)'' is to recall that this effective action is 
relevent to the weak coupling regime.}
  effective action, 
with cosmological constant $\mu_c$, background metric
$\ghat$:
\begin{equation}
S^{\rm (w)}_{\mu_c} =-{1\over 8\pi}\int dz d\zb \sqrt{-\ghat}     \{ \ghat^{ab}
\partial_a\Phi
\partial_b \Phi  +Q_L  R^{(2)} \Phi  +\mu_c e^{\alpha_- \Phi} \},
\label{1.1}
\end{equation}
The two screening charges $\alpha_\pm$ and $Q_L$ are related as usual:
\beq
\alpha_-\alpha_+=2,\quad \alpha_-+\alpha_+=Q_L,  
\label{charges}
\eeq
and the Liouville central charge is such that 
\beq
C_L=1+3Q_L^2. 
\label{central}
\eeq
For $C_L>25$, $\alpha_\pm$ are real; this is the weak coupling regime. 
For $25>C_L>1$, $\alpha_\pm$ are complex; this is the strong coupling situation 
which we want to study further in the present article. 
The Virasoro weight of the operator $\exp (-\beta \Phi)$ is
$\Delta_{\cal G}=-{1\over 2} \beta (\beta+Q_L)$.
Recall the basic point of  DDK. Consider the correlator
\beq
\left < \prod_\ell e^{-\beta_\ell \Phi(z_\ell, \zb_\ell)} \right
>^{(\nu_h)}_{\mu_c}
\equiv \int {\cal D} \phi e^{S^{\rm (w)}_{\mu_c}} \left (
\prod_\ell e^{-\beta_\ell \phi(z_\ell, \zb_\ell)}  \right )^{(\nu_h)},  
\label{1.4}
\eeq
$\nu_h$ is the number of handles.
The notation is that  $\Phi$ is  the Liouville
operator, while  $\phi$ is the c-number field that
 appears in the functional integral.
The latter is unchanged if we change variable by letting
$\phi=\widetilde \phi -\ln (\mu_c)/\alpha_-$. This gives
\beq
\left < \prod_\ell e^{-\beta_\ell \Phi(z_\ell, \zb_\ell)} \right
>^{(\nu_h)}_{\mu_c}=
\left < \prod_\ell e^{-\beta_\ell \Phi(z_\ell, \zb_\ell)} \right >^{(\nu_h)}_1
\mu_c^{[\sum _\ell\beta_\ell+Q_L(1-\nu_h)]/\alpha_-}. 
\label{1.5}
\eeq
the last term comes from the linear term of the action and
the Gauss-Bonnet
theorem
\beq
{1\over 8\pi}\int \int dz d\zb R^{(2)}=(1-\nu_h). 
\label{1.6}
\eeq
  Now, we 	
compare this procedure with the corresponding discussion in  the operator method.  
There,  the operator algebra of each chiral component has 
 a  quantum group symmetry\footnote{
For recent developments in this connection see refs.\cite{CGS1},\cite{CGS2}.}
of  the type 
$U_q(sl(2))\odot U_{\qhat}(sl(2))$, with
\beq
q=e^{ih},\quad 
\qhat=e^{i\hhat}, \qquad 
h=\pi {\alpha_-^2\over 2},\quad 
\hhat=\pi {\alpha_+^2\over 2}. 
\label{qprms}
\eeq
The two quantum group parameters are related by 
\beq
h\hhat=\pi^2,\quad h+\hhat={C_L-13\over 6}.
\label{h-hhat}
\eeq
The first relation shows that they are, in a sense, dual pairs. The above symbol $\odot$ has
 a special   meaning which was much discussed before\cite{CGR,GR1,CGS1}. 
The Hilbert space of states is of course 
 a direct sum of products of a left and a right Virasoro Verma modules. These 
 are characterized by the eigenvalue of the 
rescaled zero-mode momenta $\varpi$, $\varpib$  of the B\"acklund free field with chiral 
components $\vartheta_R(x)$, $\vartheta_L(\overline x)$ defined by writing  
on the cylinder, 
\beqa
\vartheta_R(x)&=&q_0+ip_0 x+i\sum_{n\not=0} e^{-inx} {p_n\over n},\quad
\varpi={2i\over \alpha_-}p_0, \nnn
\vartheta_L({\overline x})&=&{\overline q}_0+i{\overline p}_0 {\overline x}
+i\sum_{n\not=0} e^{-in{\overline x}} {{\overline p}_n\over n},\quad
\varpib={2i\over \alpha_-}{\overline p}_0. 
\label{p0def}
\eeqa
A right Verma module is charaterized by  the highest-weight eigenvalue 
\beq
\Delta(\varpi)={h \over 4\pi}(\varpi_0^2-\varpi^2), \quad 
\varpi_0=1+{\pi\over h}
\label{Delta}
\eeq
of $L_0$, with similar formulae for the left  Verma modules. 
For $\varpi=\pm \varpi_0$, $\Delta$ vanishes. This describes the two 
$Sl(2,C)$ invariant states.  
 In ref.\cite{G5}, it was shown that
the transformation  law Eq.\ref{1.5}
corresponds to the following  definition for  arbitrary cosmological 
constant: 
\beq
e_{(\mu_c)}^{\textstyle -\beta
\Phi}=
 \mu_c^{\beta/\alpha_-}\> \mu_c^{-\varpi/2}
e^{\textstyle -\beta\Phi}
\>\mu_c^{\varpi/2}.
\label{1.9}
\eeq 
From this it is easy to recover the standard results for  
$\nu_h=0$, and $1$. First, consider the sphere ($\nu_h=0$).  
In the operator method, the left hand side of  Eq.\ref{1.5} is given by 
\beq
\left < \prod_\ell e^{-\beta_\ell \Phi(z_\ell, \zb_\ell)} \right
>^{(0)}_{\mu_c}=
<-\varpi_0, -\varpi_0|\prod_\ell
e_{(\mu_c)}^{\textstyle -\beta_\ell \Phi}|\varpi_0, \varpi_0>.
\label{sphere} 
\eeq
where the notation $|\varpi_0, \varpi_0>$ represents the state
 with  left and right 
weights $\Delta=\overline \Delta=0$. Making use of Eq.\ref{1.9}, we 
get 
$$
\left < \prod_\ell e^{-\beta_\ell \Phi(z_\ell, \zb_\ell)} \right
>^{(0)}_{\mu_c}=
\mu_c^{\sum _\ell\beta_\ell} 
<-\varpi_0, -\varpi_0| \mu_c^{-\varpi/2} \prod_\ell
e^{\textstyle -\beta_\ell \Phi} \mu_c^{\varpi/2}
|\varpi_0, \varpi_0>.
$$
The contribution of the term $Q_LR^{(2)}\Phi$ of the effective action 
is recovered when $\mu_c^{\pm\varpi/2}$ hit the left and right vacuua, 
since, according to Eq.\ref{charges},  $Q_L/\alpha_-=\varpi_0$, and the result follows for $\nu_h=0$. 
Next, for  the
torus $\nu_h=1$, one should take the trace in the Hilbert space: 
\beq
\left < \prod_\ell e^{-\beta_\ell \Phi(z_\ell, \zb_\ell)} \right
>^{(1)}_{\mu_c}=
\hbox {\rm  Tr}\>\left [\prod_\ell
e_{(\mu_c)}^{\textstyle -\beta_\ell \Phi}\right].
\label{torus} 
\eeq
Making again use of Eq.\ref{1.9}, we now get
$$
\left < \prod_\ell e^{-\beta_\ell \Phi(z_\ell, \zb_\ell)} \right
>^{(0)}_{\mu_c}=
\mu_c^{\sum _\ell\beta_\ell}
\hbox {\rm Tr}\>\left [ \mu_c^{-\varpi/2} \prod_\ell
e^{\textstyle -\beta_\ell \Phi} \mu_c^{\varpi/2} \right ]
=\mu_c^{\sum _\ell\beta_\ell} 
\left < \prod_\ell e^{-\beta_\ell \Phi(z_\ell, \zb_\ell)} \right
>^{(0)}_{1} 
$$ 
 Thus, we get the same result without the term $Q_L/\alpha_-$. 
This agrees   with Eq.\ref{1.5} with $\nu_h=1$.

In order to  compute the string susceptibilty, we  next consider 
\beq
{\cal Z}^{(\nu_h)}_{(\mu_c)}(A)\equiv \left < \delta\left [\int dz d\zb
 e^{\textstyle \alpha_ -\Phi}-A\right ] \right >^{(\nu_h)}_{\mu_c}.  
\label{1.10}
\eeq  
It is clear, from the form of the $\mu_c$ dependence of $\exp(-\beta \Phi)$ that,  without 
entering into the detailed definition of the $\delta$ function, we should have 
\beq
\delta\left [\int dz d\zb
 e^{\textstyle \alpha_ -\Phi}-A\right ]_{\mu_c}= 
\mu_c^{-\varpi/2} \delta\left [\int dz d\zb
 \mu^{-1}_c e^{\textstyle \alpha_ -\Phi}-A\right ]_{1} 
\mu_c^{\varpi/2}. 
\label{weyldelta}
\eeq
Consider the sphere, one gets  
$$
{\cal Z}^{(0)}_{(\mu_c)}(A)=<-\varpi_0,-\varpi_0|
\mu_c^{-\varpi/2} \delta\left [\int dz d\zb
 \mu_c^{-1} e^{\textstyle \alpha_ -\Phi}-A\right ]_{1} 
\mu_c^{\varpi/2} |\varpi_0,\varpi_0>.  
$$
$$
{\cal Z}^{(0)}_{(\mu_c)}(A)=\mu_c^{1+Q/\alpha_-} {\cal Z}^{(0)}_{(1)}(A\mu_c). 
$$
Assuming that for large area ${\cal Z}^{(0)}_{(\mu_c)}(A)\sim 
A^{\gamma_{\hbox {\scriptsize   str}} -3}$, 
this gives the well known result
\beq
\gamma_{\hbox {\scriptsize   str}}=2-Q/\alpha_-=1-{\pi\over h}.
\label{1.11a}
\eeq
A similar discussion also gives back the standard result for $\nu_h=1$.

Next we describe the cosmological dependence of the 
chiral components which is such that Eq.\ref{1.9} 
holds. Our basic assumption will be   that it remains the same for the strong 
coupling regime. Contact with the quantum group classification is made by 
letting 
\beq
\beta=\alpha_- J+\alpha_+ \Jhat.
\label{1.7}
\eeq
 Then  the corresponding local
Liouville exponential operator  is given by\cite{G3,GS}
\beq
e^{\textstyle -\beta \Phi(z, \zb )}=
\sum _{m,\, \mhat}
\Vt_{m\,\mhat}^{(J,\,\Jhat)}(z)\,
\Vtb_{m\,\mhat}^{(J,\,\Jhat)}(\zb) \>
\label{Lexpdef}
\eeq
This form is dictated by locality and closure  under  fusion. The notation 
for the chiral operators refers to a particular 
normalization  whose precise expression will be  needed later on.  
In ref.\cite{GR2}, it was remarked that 
Eq.\ref{1.9} is derivable from the following 
 generalised Weyl transformation law 
\beqa
\Vt_{m\, \mhat\>\>(\mu)}^{(J\, \Jhat )} &=&   \mu^{J+\Jhat \pi/h}
\mu^{-\varpi/2} \Vt_{m\, \mhat}^{(J\, \Jhat )}
\>\mu^{\varpi/2} \nnn
\Vtb_{\mb\, \mhatb\>\>(\mub)}^{(\Jb\, \Jhatb )} &=&   \mub^{\Jb+\Jhatb \pi/h}
\mub^{-\varpib/2} \Vtb_{\mb\, \mhatb}^{(\Jb\, \Jhatb )}
\>\mub^{\varpib/2}.
\label{1.10b}
\eeqa
Taking  two different parameters $\mu$ and $\mub$ is only relevent  for 
the strong coupling case. For the
weak coupling one,  the left and right quantum numbers 
are  all equal (chirality is confined) so that in deriving Eq.\ref{1.9}, one only 
deals with the product $\mu\mub$.  The previous discussion is immediately recovered
with $\mu_c=\mu\mub$.
\subsection{The strong coupling regime}
At this point we turn to the strong coupling regime.
Now $1<C_L<25$. The screening charges
$\alpha_\pm$ are complex and related by complex conjugation.
Thus complex weights appear in general.  
The weight of a chiral operator  $\Vt_{m\, \mhat}^{(J\, \Jhat )}$ is equal to 
$\Delta(\varpi_{J,\Jhat})$, with $\varpi_{J,\Jhat}=\varpi_0+2J+2\Jhat \pi/h$. 
We shall denote by $|J,\Jhat>$ the corresponding highest weight state. 
There are  two types of
exceptional cases such that the weights are real. These are 
the states $|J,\,J>$ (resp. $|-J-1,\,J>$), with  
negative (resp. positive) weights. One could try to work with the
corresponding Liouville exponentials
$\exp[-J(\alpha_--\alpha_+)\Phi]$
(resp $\exp[\bigl((J+1)\alpha_--\alpha_+\bigr)\Phi]$), but this
would be inconsistent, since these operators do not form a closed
set under fusing and braiding. Moreover,the chiral vertex operators are such that  
\beq
<J_2,\, \Jhat_2 |  \Vt^{(J\Jhat)}_{m \mhat} | J_1,\, \Jhat_1>
\propto \delta_{J_1,\, J_2-m}\, \delta_{\Jhat_1,\, \Jhat_2-\mhat}.
\label{shift}
\eeq
As a result, it follows from  Eqs.\ref{Lexpdef}, that  Liouville exponential 
operators do not preserve the reality
condition for highest weights just recalled. The basic problem is
that Eq.\ref{Lexpdef} involves the $\Vt$ operators with arbitrary
$m$ $\mhat$, while the reality condition forces us to only use  $\Vt$
operators of the type
\beq
V_{m,\, +}^{(J)}\equiv \Vt_{-m\, m}^{(-J-1,J)},\> \hbox{or} 
\quad
V_{m,\, -}^{(J)}\equiv \Vt_{m\, m}^{(J,J)}.
\label{VJpmdef}
\eeq
 Next recall that the  truncation theorems
 hold for
\beq
C=1+6(s+2),\quad s=0,\pm 1.
\label{Cspecdef}
\eeq
For these values, there exists a closed  chiral  operator algebra  
restricted to the physical
Hilbert space
\beq
{\cal H}_{ \hbox{\footnotesize phys}}^{\pm}\equiv
   \bigoplus_{r=0}^{1\mp s}   \bigoplus_{n=-\infty}^\infty
{\cal H}^\pm_{r/2(2\mp s)+n/2}
\label{Hphysdef}
\eeq
where  ${\cal H}^\pm_{J}$  denotes the Verma modules with highest
weights $|\mp(J+1/2)-1/2,\, J>$. 
The physical operators $\chi^{(J)}_\pm$
are defined for arbitrary\footnote{
By the symbol
${\cal Z}/(2\pm s)$, we mean the set of numbers $r/(2\pm s)+n$,
with
$r=0$, $\cdots$, $1\pm s$, $n$ integer; $\cal Z$ denotes the set
of all positive or negative integers,
including zero.}  $2J \in {\cal Z}/(2\mp s)$, and
$2J_1 \in {\cal Z}/(2\mp s)$.
 to be such that\footnote{$\cal Z_+$ denotes the set of non
negative
integers.}
\beq
\chi^{(J)}_\pm\>
{\cal P}_{{\cal H}^\pm_{J_1}}
=\sum_{\nu \equiv J+m\in \cal Z_+}
(-1)^{(2\mp s)(2J_1+\nu(\nu+1)/2)}
V_{m,\, \pm}^{(J)} \>{\cal P}_{{\cal H}^\pm_{J_1}},
\label{chidef}
\eeq
where ${\cal P}_{{\cal H}^\pm_{J_1}}$ is the projector
on ${\cal H}^\pm_{J_1}$.   
According to the formulae previously recalled, their weights are, respectively,
\beq 
\Delta^-(J, C_L)=-{C_L-1\over 6}\,J(J+1), \quad
\Delta^+(J, C_L)=1+{25-C_L\over 6}\, J(J+1).
\label{2.8}
\eeq
Note that the cases of positive and negative weights are 
completely separated operatorially, eventhough we discuss them simultaneously to avoid repetitions.  
Moreover, the definition of $V_{m,\, +}^{(J)}$ is not symmetric between the two screening charges. 
For the left components, it is appropriate to make the other choice\footnote{
This is actually  a notational convenience, since clearly the two 
choices are exchanged by letting $J\to -J-1$}
\beq
\overline V_{m,\, +}^{(J)}\equiv \Vtb_{-m\, m}^{(J,-J-1)},\>  
\quad
\overline V_{m,\, -}^{(J)}\equiv \Vtb_{m\, m}^{(J,J)}.
\label{VJpmbdef}
\eeq
With this,  one defines the left physical fields by a formula similar to Eq.\ref{chidef}:
\beq
{\overline \chi}^{({\overline J})}_\pm\>
{\cal P}_{{\cal {\overline H}}^\pm_{{\overline J}_1}}
=\sum_{{\overline \nu} \equiv {\overline J}+{\overline m}\in \cal Z_+}
(-1)^{(2\mp s)(2{\overline J}_1+{\overline \nu}({\overline \nu}+1)/2)}
{\overline V}_{{\overline m},\, \pm}^{({\overline J})} \>
{\cal P}_{{\cal {\overline H}}^\pm_{{\overline J}_1}},
\label{chibdef}
\eeq
Since Eqs.\ref{2.8} are invariant under $J\to -J-1$, we get the same formuale for the 
weights of the left physical operators just defined: 
\beq 
{\overline \Delta}^-({\overline J}, C_L)=-{C_L-1\over 6}\,{\overline J}({\overline J}+1), \quad
{\overline \Delta}^+({\overline J}, C_L)=1+{25-C_L\over 6}\, {\overline J}({\overline J}+1),
\label{2.8b}
\eeq
The particular way to define the left and right components just summarised has 
another motivation, which we stressed already before. Complex conjugation exchanges 
the two screening charges, so that   the operators 
$\chi_+^{(J)}$, and ${\overline \chi}_+^{({\overline J})}$ 
are not hermitian. 
The present choice ensures that the amplitudes are nevertheless invariant under complex 
conjugation if we exchange left and right quantum numbers.
   
Now we have enough recollection to turn back to the string susceptibility. 
As is well known,   
the basic tool of the weak coupling derivation, was the existence of the 
cosmological term $\exp (\alpha_- \Phi)$, of weight $(1,1)$, whose integral over $z$ $\zb$ 
defines the invariant area and  may be added to the free field action (see Eq.\ref{1.1}) 
without breaking conformal invariance. 
In the strong coupling regime, this cosmological term 
is not acceptable since it does not preserve the reality conditions.  
On the other hand,  Eqs.\ref{2.8} \ref{2.8b} are  clearly such that  
$ \Delta^+( 0, C_L)={\overline \Delta}^+(0, C_L)=1$. As we pointed out earlier, 
the corresponding operator    
\beq
\vertex 0, 0, =\chi_+^{(0)}(z) \chib_+^{(0)}(\zb).
\label{cosmodef}
\eeq
defines a new cosmological term and  
 is local. 
Thus the area element of the strong coupling regime is
$\chi_+^{(0)}(z) \chib_+^{(0)}(\zb) dz d\zb$. It is factorized
into
a simple product of a single $z$ component by a $\zb$ component.
The effective action Eq.\ref{1.1} should be modified 
accordingly. However, it is
reasonable to assume that the behaviour under global rescaling
(global Weyl transformation) of the chiral components 
 will remain essentially the same. Moreover, the reality condition 
recalled above led us\cite{GR2} to take $\mub=\mu^{h/\pi}$, 
with $\mu$ real. 
According to Eq.\ref{1.10b}, this gives 
\beqa
\chi_{+\, (\mu)}^{(J)}&=&\mu^{-J-1+J\pi/h} \mu^{-\varpi/2}
\chi_{+}^{(J)} \mu^{\varpi/2}\nnn
\chib_{+\, (\mu)}^{(\Jb)}&=& \mu^{\Jb h/\pi -(\Jb+1)}
\mu^{-\varpib h/2\pi}
\chib_{+}^{(\Jb)} \mu^{\varpib h/2\pi}. 
\label{1.11b}
\eeqa
For the cosmological term this gives 
\beq
{\cal V}^{0,0}_{(\mu_c)}=\mu_c^{-1} \mu_c^{-(\varpi+\varpib h/\pi)/4}\> {\cal V}^{0,0}_{(1)} 
\> \mu_c^{(\varpi+\varpib h/\pi)/4}, 
\label{cosweyl}
\eeq
where we have let  $\mu_c=\mu^2$, so that the overall 
factor becomes $\mu_c^{-1}$ as it should. From there, the computation of the string susceptibility, 
already summarised in ref.\cite{GR2} goes as follows. 
  Consider, now the expectations
values similar to Eq.\ref{1.10}
\beq
{\cal Z}^{(\nu_h)}_{\mu_c}(A)\equiv \left < \delta\left [\int dz d\zb
\chi_{+\, (\mu_c)}^{(0)} \chib_{+\, (\mu_c)}^{(0)} -A\right ] \right >_{\mu_c}^{(\nu_h)}.
\label{1.16b}
\eeq 
At this time, we only have the operator method at our disposal, so that we may define 
the right hand side only for $\nu_h=0$, and $\nu_h=1$ as, respectively,  
\beqa
{\cal Z}^{(0)}_{\mu_c}(A)&\equiv&  
<-\varpi_0,\, -\varpi_0|\delta\left [\int dz d\zb
{\cal V}^{0,0} -A\right ]_{(\mu_c)} |\varpi_0,\, \varpi_0>, 
\nnn
{\cal Z}^{(1)}_{\mu_c}(A)&\equiv&  
\hbox{Tr}\left \{\delta\left [\int dz d\zb
{\cal V}^{0,0} -A\right ]_{(\mu_c)}\right \}. 
\label{Zsdef}
\eeqa
Again we shall not need the detailed definition of the delta function. Only the 
Weyl tramsformation, analogous to Eq\ref{weyldelta} will matter, that is 
\beq
\delta\left [\int dz d\zb
{\cal V}^{0,0} -A\right ]_{(\mu_c)}= 
\mu_c^{-(\varpi+\varpib h/\pi)/4}\> \delta\left [\int dz d\zb
\mu_c^{-1}{\cal V}^{0,0} -A\right ]_{(1)}\mu_c^{(\varpi+\varpib h/\pi)/4}\>. 
\label{weyldeltas}
\eeq 
From there on the calculation proceeds exactly as in the weak coupling regime, and 
we shall not repeat it . The basic point is 
that the factors $\mu_c^{\pm (\varpi+\varpib h/\pi)/4}$ give real eigenvalues when they hit 
the left or right vacua, since $\varpi_0(1+ h/\pi)=(C_L-1)/6$. This is contrast with the 
weak coupling factors $\mu_c^{\pm \varpi/2}$ that would give complex result 
here.  It came out from our choice of 
$\mu$, $\mub$, which therefore ensures the reality of the string susceptibility. 
For $\nu_h=0$ one finds $\gamma_{\hbox{str}}=(2-s)/2$. Comments about this result were given 
in ref.\cite{GR2}.  

\section{The three-point functions}
\subsection{The gravity coupling constants revisited}
The Verma modules  of gravity are conveniently characterized by the eigenvalue 
of the rescaled momentum $\varpi$. The spectrum of eigenvalue is of the 
form $\varpi_{J,\, \Jhat}=\varpi_0+2J+2\Jhat\pi/h$.  The starting point to
derive the three-point functions is the expression\cite{GR1} of the 
matrix element of the $\Vt$ fields
recalled above between such highest weight states. It is convenient to 
rewrite it under the form 
$$ <-\varpi_3| \Vt^{(J_1 \Jhat_1 )}_{m \mhat}| \varpi_2> =
\delta_{J_2-J_3-m,0}\, \delta_{\Jhat_2-\Jhat_3-\mhat,0}
g_{\varpi_1,\varpi_2}^{-\varpi_{3}}
$$
\beq
g_{\varpi_1,\varpi_2}^{-\varpi_{3}}= (-1)^{\nu \nuhat}
(i/2)^{\nu+\nuhat}{H_{\nu \nuhat}(\varpi_1)H_{\nu\nuhat}(\varpi_2)
H_{\nu\nuhat}(\varpi_{3})
\over{H_{\nu\nuhat}(\varpi_{\nu/2,\nuhat /2})}}
\label{ggen}
\eeq
with
\beqa
\varpi_i&=&\varpi_0+2J_i+2{\pi\over h} \Jhat_i, \quad i=1,\, 2, \, 3; \nnn 
 \nu+\nuhat \pish&=&\demi (\varpi_1+\varpi_2+\varpi_{3}-\varpi_0)\equiv \nue
\label{pdef}
\eeqa
The $g$'s are called coupling constants. 
Each factor $H_{\nu\nuhat}(\varpi)$ was shown to be expressible as a sum
over path. It is convenient to write it 
 under the form (with $\varpi=\varpi_0+2J+2\Jhat \pish$)
\beq
H_{\nu\nuhat}(\varpi)=\left ({-\pi\over h}\right)^{-(1/4+\Jhat)\nu} 
\left ({-h\over \pi}\right)^{-(1/4+J)\nuhat} \chemin \varpi-\nue,{\varpi} .
\label{chemdef}
\eeq
In general we  define  the symbol $\chemin A^e ,{B^e} $ for $B^e-A^e\equiv
\nue=\nu+\nuhat \pish$, with
$\nu$ and $\nuhat$ positive integers, as the following 
 product of  factors 
along a general path
$$
\chemin A^e,{B^e} =\prod_{\ell=0}^{N-1}
\left\{\left(-\pish\right)^{\Rhat_\ell-1/2}
 F\left[{\Rhat^e_\ell+\Rhat^e_{\ell+1}\over 2}-{h\over 2\pi}
\right] \right\}^{\epsilon_\ell/2} \times
$$
\beq
\prod_{\ell=0}^{N-1}
\left\{ \left(-\hspi\right)^{R_\ell-1/2}
 F\left[{R^e_\ell+R^e_{\ell+1}\over 2}-{\pi\over 2h}
\right] \right\}^{\epsilonhat_\ell/2},
\label{chemdef1}
\eeq
where the product is taken along any path going from $A^e$ to $B^e$ with
intermediate  points
$R_\ell^e$, such that   
$$
R_{\ell+1}^e-R_\ell^e=\epsilon_\ell+\pish \epsilonhat_\ell, \quad
\epsilon_\ell=0,\, \pm 1, \>
\epsilonhat_\ell=0,\, \pm 1,\quad \epsilon_\ell \epsilonhat_\ell=0;
$$
\beq
R_0^e=A^e,\quad R_{N}^e=B^e, \quad \Rhat_\ell^e=R^e_\ell \hspi, 
\quad R_\ell^e=R_\ell+\pish \Rhat_\ell.
\label{chemdef2}
\eeq
Note that we shall always be in  the case where  
$R_\ell$ and $\Rhat_\ell$ are rational, while $\hspi$ is not. Thus the last equation 
is unambiguous.  We shall not bother about the precise
choice of sheet for square root of gamma functions. 
It should be specified according to  
ref.\cite{G5}. It is not important since it will only appear in the 
final leg factors. This product does not depend 
upon the choice of
path\footnote{
This may be seen, in particular, by making use of ref.\cite{G}}.
 Up to the factor we have put in front of 
Eq.\ref{chemdef} it coincides with the one introduced in 
refs.\cite{GR1}, \cite{CGR}. 
One may visualise the path as made up with $N$ segments of coordinates
$\epsilon_\ell$,
$\epsilonhat_\ell$,
starting from the point specified by $R^e_\ell$. 
 Using the fact that $\epsilon_\ell\epsilonhat_\ell=0$, one may 
rewrite the factors  $\left(-\pi/h\right)^{\Rhat_\ell-1/2}$, 
and 
$ \left(-h/\pi\right)^{R_\ell-1/2}$ may be rewritten, respectively as 
$\left(-\pi/h\right)^{(\Rhat_\ell+\Rhat_{\ell+1}-1)/2}$, 
and 
$ \left(-h/\pi\right)^{(R_\ell+R_{\ell+1}-1)/2}$, so that our definition 
only involves the midpoints $(R^e_\ell+R^e_{\ell+1})/2$ of each segment. It is 
easy to verify that we have the composition law 
\beq
\chemin A^e,{B^e} \>  \chemin B^e,{C^e} =\chemin A^e,{C^e}
\label{mult}
\eeq
So far, we assumed that $\nu$ and $\nuhat$ are positive integers. This last
relation allows us to
extend the definition to arbitrary signs. Next, it is useful to use the fact that 
$F(z)$ satisfies the relation $F(1-z)=(F(z))^{-1}$ to derive the
 identity
\beq
\chemin A^e,B^e =\chemin 1+\pish-A^e,{1+\pish-B^e} . 
\label{sym}
\eeq
Returning to Eq.\ref{ggen}, one finds that it may be written as 
\beq
g_{\varpi_1,\varpi_2}^{-\varpi_{3}}= (-1)^{\nu\nuhat} \left({i\over 2}\right)^{\nu+\nuhat} 
\left(-{\pi\over h}\right)^{(\nu-\nuhat)/2}
\chemin \varpi_1-\nu^e,{\varpi_1}
\chemin \varpi_2-\nu^e,{\varpi_2}
\chemin \varpi_3-\nu^e,{\varpi_3}
\chemin {-\nue},{0} .
\label{gchem}
\eeq
The last term has been retransformed using Eq.\ref{sym} so that it takes
the same form
as the other three with $\varpi=0$. Letting $\varpi_4=0$, we may thus
write compactly
\beq
g_{\varpi_1,\varpi_2}^{-\varpi_{3}}= (-1)^{\nu\nuhat} \left({i\over 2}\right)^{\nu+\nuhat} 
\left(-{\pi\over h}\right)^{(\nu-\nuhat)/2}
\prod_{i=1}^4  \chemin \varpi_i-\nu^e,{\varpi_i}
\label{gcomp}
\eeq 
\subsection{The dressings}
\subsubsection{The weak coupling case} 
In the weak coupling regime, one  represents\cite{G5} matter by another copy of the
Liouville theory  with a different central charge. Following our previous 
works the symbols pertaining to matter are  noted with a prime 
(or, if convenient, with an index $M$). The relations between matter and gravity parameters 
 are 
\beq
C_M=26-C_L, \quad h/\pi=-\pi/h',\quad \alpha'_\pm=\mp i\alpha_\mp .
\label{basic}
\eeq  
One constructs local
fields in analogy with Eq.\ref{Lexpdef}:
\beq
e^{\textstyle -(J'\alpha'_-+\Jhat'\alpha'_+)\Phi'(z, \zb )}=
\sum _{m,\, \mhat}
\Vt'\,^{(J',\,\Jhat')}_{m\,\mhat}(z)\,
\Vtb'\,^{(J',\,\Jhat')}_{m\,\mhat}(\zb).  \>
\label{L'expdef}
\eeq
 $\Phi'(z, \zb )$ is the matter field  (it
commutes with $\Phi(z, \zb )$),
There are  two possible  dressing of these operators by gravity  such 
that the total weights are $\Delta=\bar \Delta=1$. 
The first  is
achieved by considering the vertex
operators
\beq
{\cal W}^{J',\Jhat'}\equiv
e^{\textstyle -((-\Jhat'-1)\alpha_-+J'\alpha_+)\Phi
 -(J'\alpha'_-+\Jhat'\alpha'_+)\Phi'}
\label{vertexdef}
\eeq
In particular for $J'=\Jhat'=0$, we get the cosmological term
$\exp (\alpha_-\Phi)$. For the associated Liouville zero modes, this 
means that $\varpi_{J, \Jhat}=-\varpihat'_{J',\Jhat'}$.   
The other choice of dressing leads to the operators
\beq
\vertexcw {J'},{\Jhat'}, 
=
e^{\textstyle -(\Jhat'\alpha_--(J'+1)\alpha_+)\Phi
 -(J'\alpha'_-+\Jhat'\alpha'_+)\Phi'}. 
\label{vcdef}
\eeq
\subsubsection{The strong coupling case}
We consider another copy of the 
strongly coupled theory, with central charge $C_M=26-C_L$. Since
this gives $C_M=1+6(-s+2)$, we are also at the special values, and
the truncation theorems applies to matter as well. This ``string
theory'' has no transverse degree of  freedom, and is thus
topological.
The complete dressed vertex operator are  now  
\beq
\vertex {J'}, {\Jb'},
=
\chi_+^{(J')} \chib_+^{(\Jb')}\>
\chi'\, ^{(J')}_-  \chib'\, ^{(\Jb')}_-, 
\label{VertexSdef}
\eeq 
\beq
\vertexc {J'}, {\Jb'},
=
\chi_+^{(-J'-1)}  \> \chib_+^{(-\Jb'-1)} \>
\chi'\,^{J')}_- \>\chib'\,^{\Jb'}_-.
\eeq
As in the weak coupling formula, operators relative to
matter are distinguished by a prime. The gravity  part is taken to have positive weights 
so that it includes the cosmological term.

\subsection{The matter  coupling constants revisited} 
It is convenient to   define  symbols $\cheminp A'{}^e,B'{}^e $ related to matter, 
similar to  $\chemin A^e,B^e $. Let   
$$
\cheminp A'{}^e,{B'{}^e} =\prod_{\ell=0}^{N-1}
\left\{\left(\pishp\right)^{\Rhat'_\ell-1/2}
 F\left[{\Rhat'{}^e_\ell+\Rhat'{}^e_{\ell+1}\over 2}-{h'\over 2\pi}
\right] \right\}^{\epsilon'_\ell/2} \times
$$
\beq
\prod_{\ell=0}^{N-1}
\left\{ \left(\hspip\right)^{R'_\ell-1/2}
 F\left[{R'{}^e_\ell+R'{}^e_{\ell+1}\over 2}-{\pi\over 2h'}
\right] \right\}^{\epsilonhat'_\ell/2}. 
\label{chemdefp}
\eeq
 It is the direct analogue of Eq.\ref{chemdef2}, apart from a simple 
modification, namely, we  put the factors  $\left(\pi/h'\right)^{\Rhat'_\ell-1/2}$, 
and 
$ \left(h'/\pi\right)^{R'_\ell-1/2}$, instead of  
$\left(-\pi/h'\right)^{\Rhat'_\ell-1/2}$, 
and 
$ \left(-h'/\pi\right)^{R'_\ell-1/2}$.  
  As we alradry know\cite{G5,GR1,GR2}, tremendous simplifications occur when gravity
and matter are
put together. Our definitions are such that this 
appears neatly 
the level of
the product symbols. Indeed, let us derive the following basic relations 
\beq
\cheminp {A'{}^e},{B'{}^e} = \chemin {1+\pish B'{}^e},{1+\pish
A'{}^e} =
 \chemin {\pish(1- B'{}^e)},{\pish(1- A'{}^e)} . 
\label{basescreen}
\eeq
The equality between the   last two expressions is a direct consequence of Eq.\ref{sym}. 
In order to derive the first equality, one  transforms the first line of 
Eq.\ref{chemdefp} by writing 
$$F\left[{\Rhat'{}^e_\ell+\Rhat'{}^e_{\ell+1}\over 2}-{h'\over 2\pi}
\right]=\left\{ F\left[1-{\Rhat'{}^e_\ell+\Rhat'{}^e_{\ell+1}\over 2}-{\pi\over 2h}
\right]\right\}^{-1}
$$
Let us define  the associated path in Liouville theory by 
$R_\ell^e=1-\Rhat'{}^e_\ell$, so that the second term becomes   the 
inverse of the Liouville factor  $F\left[{R^e_\ell+R^e_{\ell+1}\over 2}-{\pi\over 2h}
\right]$. On the other hand, an easy calculation shows that  for the other $F$ we have 
$$
F\left[{R'{}^e_\ell+R'{}^e_{\ell+1}\over 2}-{\pi\over 2h'}\right]= 
F\left[{\Rhat{}^e_\ell+\Rhat{}^e_{\ell+1}\over 2}-{h\over 2\pi}\right]. 
$$
It follows from the relations given above that $R_\ell=1-\Rhat'_\ell$, and 
$\Rhat_\ell=R'_\ell$. Therefore the associated path of gravity is such that 
$\epsilon_\ell=\epsilonhat'_\ell$, $\epsilonhat_\ell=-\epsilon'_\ell$. 
This allows to verify that all the other factors also match, and Eq.\ref{basescreen} 
follows. \qed 

 Finally, the  matter coupling is given by 
\beq
g'{}_{\varpi'_1,\varpi'_2}^{-\varpi'_{3}}= \left({i\over 2}\right)^{\nu'+\nuhat'} 
\left({\pi\over h'}\right)^{(\nu'-\nuhat')/2}
\prod_{i=1}^4  \cheminp \varpi'_i-\nu'^e,{\varpi'_i}
\label{gcompp}
\eeq
\subsection{The leg factors}
In this part, we rediscuss the simplifications that occur when gravity and matter are 
put together at the most basic level, that is, when the coupling constants are multiplied. 
 As originally discussed in \cite{G5}, and as we just saw, the dressing conditions have a
simple expression
in terms of the $\varpi$'s, namely, one must have $\varpi'=\pm
\varpihat$. The two choices
of sign corresponds to the two possible screening charges, and are
treated much in the
same way. There are in fact only two really different situations, the
first when one
takes the same dressing condition on all three legs---this is the case
we have considered
in our previous works---and the second when one of the screening
charges is of the
opposite type. Let us discuss them in turn.
\subsubsection{The same choice for every legs}
This situation  was discussed at length before. We return to it  briefly
as a preparation for the
mixed screening choice.
First consider the case where   $\varpi_i'=\varpihat_i\equiv \hspi
\varpi_i$.
Note that, since by definition $\varpi_4=0$, the same screening
condition  holds (trivially)
for $i=4$. Applying the formulae given above one finds that $\nue'=\hspi
(\nue +1)$, and
$$
\cheminp {\varpi_i'-\nue'},{\varpi_i'} = \chemin {\varpi_i+1},{\varpi_i-\nue} .
$$
Making use of Eq.\ref{mult} this leads to 
$$
\cheminp {\varpi_i'-\nue'},{\varpi_i'}  \chemin {\varpi_i-\nue},{\varpi_i,} =
 \chemin {\varpi_i+1 },{\varpi_i} =
\left \{\left(-\pish\right)^{2\Jhat_i+1/2}F[\varpihat_i]\right\}^{-1/2}.
$$
Altogether, we get a product of leg factors
\beq
g_{\varpi_1 \varpi_2}^{-\varpi_3} g'{}_{\varpihat_1
\varpihat_2}^{-\varpihat_3}=(-1)^{\nu\nuhat} 
\left({i\over 2}\right)^{2\nuhat-1} \left(-\hspi\right)^{2\nu} 
\prod_{i=1}^4  {1\over \sqrt{F[\varpihat_i]}}  .
\label{meme-}
\eeq
The other screening condition $\varpi_i'=-\hspi \varpi_i$ is treated
similarly, obtaining
\beq
g_{\varpi_1 \varpi_2}^{-\varpi_3} g'{}_{-\varpihat_1
-\varpihat_2}^{\varpihat_3}=(-1)^{\nu\nuhat}
\left({i\over 2}\right)^{2\nu-1} \left(-\pish\right)^{2\nu} 
\prod_{i=1}^4  {1\over \sqrt{F[\varpi_i]}}    .
\label{meme+}
\eeq
\subsubsection{The mixed case}
The coupling constant $g_{\varpi_1 \varpi_2}^{-\varpi_3}$ is completely
symmetric and,
by relabelling,  we may always choose the third leg to be the one which
differs from the others.
Again there are two cases. First choose $\varpi_i'=\hspi \varpi_i$ for
$i=1,\, 2$,  and
$\varpi_3'=-\hspi \varpi_3$. The relation between $\nu^e$ and $\nue'$ may
be written as
 $\nue'=\hspi ( \nue-\varpi_3+1)$. A simple modification of a calculation
performed above
then gives, for $i=1,\,2$,
$$
\cheminp {\varpi_i'-\nue'},{\varpi_i'}  \chemin {\varpi_i-\nue},{\varpi_i} =
 \chemin {\varpi_i+1},{\varpi_i-\nue+\varpi_3} \chemin {\varpi_i-\nue},{\varpi_i}    
$$
\beq
\phantom{
\cheminp {\varpi_i'},{\varpi_i'-\nue'} \chemin {\varpi_i},{\varpi_i-\nue}  }
= \chemin {\varpi_i+1 },{\varpi_i}
\chemin {\varpi_i-\nue},{\varpi_i-\nue+\varpi_3}   
\label{mix1}
\eeq
where we have used the multiplication law Eq.\ref{mult}. For $i=3$, we
write instead
$\nue'=-\hspi \nue -1+\hspi (\varpi_1+\varpi_2)$. One now gets
\beq
\cheminp \varpi_3',{\varpi_3'-\nue'} \chemin \varpi_3,{\varpi_3-\nue} =
 \chemin {\varpi_3+\pish },{\varpi_3}
\chemin {\varpi_3-\nue},{\varpi_3-\nue+\varpi_1+\varpi_2}.
\label{mix2}
\eeq
The last term in Eq.\ref{gchem} may be treated in the same way as the
third,
and the result is given by
the last equation with $\varpi_3\to 0$. Altogether, one  gets
\beq
g_{\varpi_1 \varpi_2}^{-\varpi_3} g'{}_{\varpihat_1
\varpihat_2}^{\varpihat_3}= 
 \left({i\over 2}\right)^{2(J_3-\Jhat_1-\Jhat_2)+1} \left(-\pish\right)^{\nu-\nuhat} 
 {(-1)^{\nu\nuhat} \over \sqrt{F[0]}} {1\over \sqrt{F[\varpihat_1]}} 
{1\over \sqrt{F[\varpihat_2]}} {1\over \sqrt{F[\varpi_3]}}   .
\label{mix+}
\eeq
When one performes the computation, one first arrives at the right hand side 
 multiplied by the factor
$$
\chemin {\varpi_1-\nue},{\varpi_1+\varpi_3-\nue}  \chemin
{\varpi_2-\nue},{\varpi_2+\varpi_3-\nue}
\chemin {\varpi_3-\nue},{\varpi_1+\varpi_2+\varpi_3-\nue}  \chemin
{-\nue},{\varpi_1+\varpi_2-\nue}  .
$$
Making use of Eq.\ref{basescreen}, one may verify that it is equal to
one, thereby completing the derivation. Thus in this
case also the result is a product of leg factors. A similar calculation
allows us to
deal with the other choice, obtaining
\beq
g_{\varpi_1 \varpi_2}^{-\varpi_3} g'{}_{-\varpihat_1 -\varpihat_2}^{-
\varpihat_3}=
 \left({i\over 2}\right)^{2(\Jhat_3-J_1-J_2)+1} \left(-\hspi\right)^{\nuhat-\nu} 
 {(-1)^{\nu\nuhat} \over \sqrt{F[0]}} {1\over \sqrt{F[\varpi_1]}} 
{1\over \sqrt{F[\varpi_2]}} {1\over \sqrt{F[\varpihat_3]}} .
\label{mix-}
\eeq
\subsection{The three point functions.} 

The computation makes use of the    relations just derived for the coupling constants. 
  
Let us first consider the weak coupling case for comparison. One gets with the same screening 
on every legs\footnote
{up to separate multiplicative factors on each leg which we drop. See 
refs.\cite{G5}, \cite{CGR} for detailed computations of these 
factors, in  some cases. }
:
\beq
\left < \prod_\ell {\cal W}^{J'_\ell, \Jhat'_\ell}\right >=
\left (g_{-\varpihat'_1 -\varpihat'_2}^{\varpihat'_3} g'{}_{\varpi'_1 \varpi_2'}^{ 
-\varpi'_3}\right)^2= 
{1\over F[0]} \prod_{i=1}^3{1\over  F[\varpi'_i]}
\label{threepw}
\eeq
\beq
\left < \prod_{\ell=1}^3 \vertexcw {J'_\ell},{\Jhat'_\ell},  \right >=
\left (g_{\varpihat'_1 \varpihat'_2}^{-\varpihat'_3} g'{}_{\varpi'_1 \varpi'_2}^{
-\varpi'_3}\right)^2= 
{1\over F[0]} \prod_{i=1}^3{1\over  F[\varpihat'_i]}. 
\label{threepwc}
\eeq 
In the weak coupling regime with one different screening, one gets 
\beq
\left <  \vertexcw {J'_3},{\Jhat'_3}, \prod_{\ell=1}^2 {\cal W}^{J'_\ell, \Jhat'_\ell}\right >=
\left (g_{-\varpihat'_1 -\varpihat'_2}^{-\varpihat'_3} g'{}_{\varpi'_1 \varpi'_2}^{
-\varpi'_3}\right)^2= 
{1\over F[0]} {1\over  F[\varpihat'_3]}\prod_{i=1}^2{1\over  F[\varpi'_\ell]}. 
\label{threepwm}
\eeq

Next consider the strong coupling case.  First, in order to make contact with our 
previous work\cite{GR1} let us consider the completely symmetric case 
\beq
\left < \prod_\ell {\cal V}^{J'_\ell, \Jb'_\ell}\right >=
g_{-\varpihat'_1 -\varpihat'_2}^{\varpihat'_3}\> g'{}_{\varpi'_1 \varpi_2'}^{ 
-\varpi'_3}\>
g_{{\widehat {\overline \varpi}}'_1 {\widehat {\overline \varpi}}'_2}^
{-{\widehat {\overline \varpi}}'_3}\>  g'{}_{\varpib'_1 \varpib_2'}^{ 
-\varpib'_3}
=
{1\over F[0]} \prod_{i=1}^3{1\over  \sqrt{ F[\varpi'_i] F[\varpihatb'_i]}}.  
\label{threeps}
\eeq 
 Next we shall  rather make use of the non symmetric three-point function
$$
\left <  \vertexc {J'_3},{\Jb'_3}, 
\prod_{\ell=1}^2 {\cal V}^{J'_\ell, \Jb'_\ell}\right >=
g_{-\varpihat'_1 -\varpihat'_2}^{-\varpihat'_3}\> g'{}_{\varpi'_1 \varpi_2'}^{ 
-\varpi'_3}\>
g_{{\widehat {\overline \varpi}}'_1 {\widehat {\overline \varpi}}'_2}^
{{\widehat {\overline \varpi}}'_3}\>  g'{}_{\varpib'_1 \varpib_2'}^{ 
-\varpib'_3}=
$$
\beq
{1\over F[0]} {1\over  \sqrt{ F[\varpihat'_3] F[\varpib'_3]}} 
\prod_{i=1}^2{1\over  \sqrt{ F[\varpi'_i] F[\varpihatb'_i]}}.  
\label{threepm}
\eeq 
 All the expressions of coupling constants
we have obtained involve a factor $1/F(0)$ which is divergent. We drop it 
from now on, since it may be reabsorbed by an overall change of the normalization 
of the three point functions which will not matter for the forthcoming discussion. 
\section{N-point functions}
\subsection{Strong coupling regime}

As shown in ref.\cite{DFK}, and as we will recall below, the key tools in deriving the 
higher point functions of the weak coupling regime are the Ward identities which may 
be derived from the path integral formulation Eq.\ref{1.4}, by taking derivatives with respect 
to the cosmological constant $\mu_c$. In so doing, the details of the effective action 
are not important. One only uses the fact that the dependence upon this 
parameter in Eq.\ref{1.4} is entirely contained---by definition---in the last term 
of $S_{\mu_c}^{\rm (w)}$. In the strong coupling regime another cosmological term 
(${\cal V}^{0,0}$) appears, and it is reasonable to assume that the 
higher point functions will be determined from the associated ward identities. In this case, 
however, the vertex operators create and destroy gravity chirality, so that the effective 
action cannot be constructed out of an ordinary world sheet boson.  It should rather 
involve two independent chiral bosons with opposite chiralities. We leave its derivation for
further studies, since its explicit form does not matter so  much at present. 
We only assume that this theory can be described by an effective action 
including the new cosmological term:
\beq
S_{\mu_c}=S_0+\mu_c \int \vertex 0, 0,
\label{nS}
\eeq
where the action $S_0$ does not depend on the cosmological
constant $\mu_c$.
So, we consider N-point correlators
\beq
\int
e^{-S_{\mu_c}}
\vertexc {-J_1-1},{-\Jb_1-1},
\vertex J_2,\Jb_2,
...
\vertex J_N,\Jb_N, 
=
\left<
\vertexc {-J_1-1},{-\Jb_1-1},
\vertex J_2,\Jb_2,
...
\vertex J_N,\Jb_N, 
\right>_{\mu_c}
\eeq
where the integral sign stands for the functional integration
and the integration over the position of the $N$ insertion points. For notational simplicity 
we drop the primes of the spins appearing in the vertex operators. As before, 
making use of Eq.\ref{1.11b} one derives  the scaling law of these correlators:
\beq
\left<
\vertexc {-J_1-1},{-\Jb_1-1},
...
\vertex J_N,\Jb_N, 
\right>_{\mu_c}=
\left<
\vertexc {-J_1-1},{-\Jb_1-1},
...
\vertex J_N,\Jb_N, 
\right>_{1}
\>
\mu_c^{\sum_{i=1}^NP_i
-\left({N\over2}-1\right)
\left({s+2\over2}\right)}
\label{mudep3}
\eeq
with
\beq
P_i=P(J_i,\Jb_i)=
{Q_M\over 4}
[\alpha'_-(J_i+{1\over 2})
+\alpha'_+(\Jb_i+{1\over 2})]
\label{Pi}.
\eeq
Derivation with respect to the cosmological constant
of these $N$ point functions gives $N+1$ point functions with
one external momentum put to zero:
\beq
{\partial\over\partial{\mu_c}}
\left<
\vertexc {-J_1-1},{-\Jb_1-1},
\vertex J_2,\Jb_2,
...
\vertex J_N,\Jb_N, 
\right>_{\mu_c}
=
\left<
\vertexc {-J_1-1},{-\Jb_1-1},
\vertex J_2,\Jb_2,
...
\vertex J_N,\Jb_N, 
\vertex 0,0,
\right>_{\mu_c}
\label{der1}.
\eeq
If this model can indeed be described by such an effective action,
its perturbative expansion must yield Feynman rules.
We already computed the three point vertices previously
and we represent it graphically with an outgoing momentum
for the leg $(J_1,\Jb_1)$:
\beq
\left<
\vertexc {-J_1-1},{-\Jb_1-1},
\vertex J_2,\Jb_2,
\vertex J_3,\Jb_3, 
\right>_{\mu_c}
\!\!
=
\lfc J_1, \Jb_1,
\lf J_2, \Jb_2,
\lf J_3, \Jb_3,
\ 
\mu_c^{P_1+P_2 + P_3
-{s+2\over 4}}
=
\epsfbox{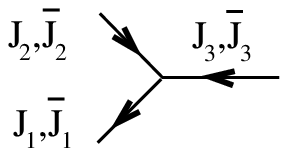}
\label{f3p2b}
\eeq
with
\beq
\lf J, \Jb, = 
{1\over
\sqrt{
F\left[ (1-\hspi)(1+2J)\right]
F(\left[ (1-\pish)(1+2\Jb)\right])
}}
,
\eeq
\beq
\lfc J, \Jb,
=
{1\over
\sqrt{
F\left[ -(1-\pish)(1+2J)\right]
F(\left[ -(1-\hspi)(1+2\Jb)\right])
}}
,
\eeq
from previous sections. The particular three-point function
\beq
\left <
\vertexc -J_i-1, -\Jb_i-1,
\vertex 0, 0,
\vertex J_i,  \Jb_i,
\right >
_{\mu_c}
=
\mu_c
^{2P_i
-1
}
\lfc J_i,\Jb_i,
\lf 0,0,
\lf J_i,\Jb_i,
\label{f3p0}
\eeq
is the derivative of the two point function.
This allows us  to compute it by taking the primitive  and get
\beq
\left <
\vertexc -J_i-1,  -\Jb_i-1,
\vertex J_i, \Jb_i,
\right >_{\mu_c}
\!\!\!\!
=
{
1
\over 2P_i}
\>
\mu_c
^{2P_i}
\>
\lfc J_i,\Jb_i,
\lf J_i,\Jb_i,
\eeq
by using that $\lf 0,0,=1$,
which can be checked for every special
value\footnote{This property is again very specific
to these central charges.}
$C_L=7,13,19$.
For amplitudes with truncated external legs,
such a two point function is the inverse of the propagator,
and its
value is consequently given by\footnote{We rescaled
by an unexplained factor $(1/2)$ which will turn out necessary
so that the Feynman rules match with the derivation ones.}
\beq
P_{\mu_c}(J_i,\Jb_i)=
P_i
\>
{
\mu_c
^{-2P_i}
\over(
\lfc J_i,\Jb_i,
\lf J_i,\Jb_i,
)}
=
\epsfbox{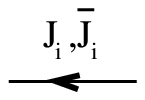}
\label{prop}.
\eeq
On the other hand, by taking derivatives, one gets higher order correlators.
The derivatives of the three point functions
yields four point functions with one external moment put to zero.
However, the $N$-point functions should be symmetric in their $N$
 legs  when expressed in terms of the variables we are using. 
 (this is true for $N=3$)
and we get  by symmetrisation the generic four-point function
$$
\left <
\vertexc -J_1-1, -\Jb_1-1,
\vertex J_2, \Jb_2,
\vertex J_3, \Jb_3,
\vertex J_4, \Jb_4,
\right >_{\mu_c}
=
\lfc J_1,\Jb_1,
\lf J_2,\Jb_2,
\lf J_3,\Jb_3,
\lf J_4,\Jb_4,
\mu_c^{P_1+P_2+P_3+P_4
-{s+2\over 2}}
$$
\beq
\left(
{Q_M\over 4}\Bigl[ \alpha'_-
(J_1+J_2+J_3+J_4+1)
+\alpha'_+(\Jb_1+\Jb_2+\Jb_3+\Jb_4+1)\Bigr]
-1
\right)
\label{f4p1}
\eeq
If we believe in this effective action description,
it must yield Feynman rules by perturbative expansion.
This total four point function \ref{f4p1} can therefore be obtained 
from the following diagrams
\beq
\epsfbox{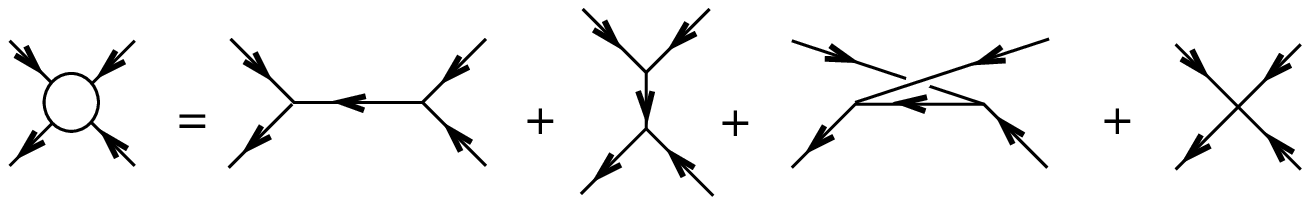}
\label{f4ps}
\eeq
the last one being the one particule irreducible amplitude. It is clear on this example 
that, by choosing one incoming and $N-1$ outgoing momenta, we have 
consistently restricted 
ourselves to  Feynman rules which 
only involve one particle irreducible vertices of the same type. 
Our purpose here is to get those one particule irreducible vertices,
which are the basic ingredients of the theory.
To do this, we shall substract from the total four point function \ref{f4p1}
the first three diagrams of \ref{f4ps} which we shall compute
from the previous Feynman rules (the propagator \ref{prop} and
the vertices \ref{f3p2b} with their selection rules\footnote{
Of the type $J_1+J_2-J_{12}$ integer...}).

Let us figure out what shall give such a computation.
In a diagram of the type
\beq
\epsfbox{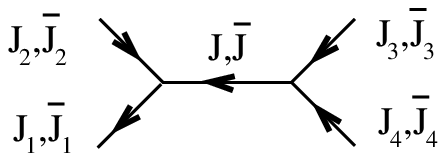}
\label{f4p1d},
\eeq
and in any diagram,
it can easily be verified that the leg factors coming from the propagators
and the ones coming from the vertices cancel out.
There only remains the leg factors attached to the external legs
and we will consequently stop writing them down\footnote{
We shall moreover only write formulae for $\mu_c=1$
as the general dependency in $\mu_c=1$ is now well known.}.
So, the value of such a diagram is essentially given by the momentum
$P(J,\Jb)$ given in \ref{Pi} coming from the propagator
\ref{prop} for $J_i=J, \Jb_i=\Jb$.
We can then easily imagine that the sum of three such diagrams
(see \ref{f4ps}) can give something not very different
from the total four point function \ref{f4p1},
and hence by difference a quite simple four point
one particule irreducible function.

However, it does not work this way.
Due to the screening charges,
or equivalently to the selection rules of the three point functions,
the momentum is not conserved.
In terms of spins, it means that in the previous diagram
one can have $J=J_3+J_4-\ns_2$ and $\Jb=\Jb_3+\Jb_4-\nsb_2$
where $\ns_2$ and $\nsb_2$ are the number of screening charges
attached to the second vertex (right hand side).
These numbers can be any positive integers provided
they do not exceed the total number of screening charges
$\ns=J_2+J_3+J_4-J_1,\nsb=\Jb_2+\Jb_3+\Jb_4-\Jb_1$
(this stems from the selection rules for the first vertex).
And so, the only consistent way to proceed is to sum over
all the possible internal momenta or spins ($J,\Jb$).
So, the value of the diagram \ref{f4p1d} will be
\beq
\sum_{\ns_2=0}^\ns
\sum_{\nsb_2=0}^\nsb
{Q_M\over 4}
[\alpha'_-(J_3+J_4-\ns+{1\over 2})
+\alpha'_+(\Jb_3+\Jb_4-\nsb+{1\over 2})]
\label{calcfeyn}.
\eeq
We notice that
these selection rules give arithmetic progressions
for the spins $J$ and $\Jb$
and that the
propagator is linear in these spins.
It is hence clear that such a sum shall yield an overall (combinatorial) factor
equal to the number of terms in it ($(\ns+1)(\nsb+1)$ in this case).
It can be verified on each case.
As a consequence the summation of the first three diagrams of \ref{f4ps}
over their possible internal momenta
will not give something close to the value of the
total four point function \ref{f4p1},
as they include this extra combinatorial factor.
We then conclude that the right value of the total four point function
should include this factor.
We consequently define the physical $N$ point amplitudes by
$$
\>^{\hbox{\scriptsize Phys}}\!
A^{(N)}_{(\ns,\nsb)}
\left(
(J_1,\Jb_1),
(J_2,\Jb_2)...
(J_N,\Jb_N)
\right)
$$
\beq
=
\left(
\,^{\ns+N-3}_{N-3}
\right)
\left(
\,^{\nsb+N-3}_{N-3}
\right)
\left<
\vertexc J_1,\Jb_1,
\vertex J_2,\Jb_2,
...
\vertex J_N,\Jb_N,
\right>.
\label{Aphys}
\eeq
The combinatorial factors of the type $\left(
\,^{\ns+N-3}_{N-3}
\right)
$
are the number of ways of placing $\ns$ screening charges
on the $N-2$ vertices of a $N$ point function\footnote{
This can be obtained by developping in $x$ the expression
$(1+x+x^2...)^{N-2}=1/(1-x)^{N-2}$.
Its coefficient of order $\nu$ is $(N-2)(N-1)...(N+\ns-3)/\ns!$,
which is equal to the number of choices $\left(
\,^{\ns+N-3}_{N-3}
\right)$.
It is remarquable that it may be expressed as a number of choices
of $N-3$ (or $\ns$) objects among $\ns+N-3$,
however we have no interpretation of it by now.}.

Of course,
this modifies the derivation rule \ref{der1} in the following way:
$$
\>^{\hbox{\scriptsize Phys}}\!
A^{(N+1)}_{(\ns,\nsb)}
\left(
(J_1,\Jb_1),
(J_2,\Jb_2)...
(J_N,\Jb_N),
(0,0)
\right)
$$
\beq
=
\left({\ns+N-2\over N-2}\right)
\left({\nsb+N-2\over N-2}\right)
\>{\partial\over\partial{\mu_c}}
\>^{\hbox{\scriptsize Phys}}\!
A^{(N)}_{(\ns,\nsb)}
\left(
(J_1,\Jb_1),
(J_2,\Jb_2)...
(J_N,\Jb_N)
\right)
\label{der2},
\eeq
which gives now for the four point function\footnote{
It is easy to see that the new derivation rule \ref{der2}
does not change anything to our previous computation of the two point
function from the three point one,
as the number of screenings is zero in that case.}
$$
\epsfbox{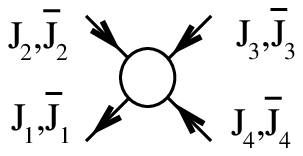} =
\>^{\hbox{\scriptsize Phys}}\!
A^{(4)}_{(\ns,\nsb)}
\left(
(J_1,\Jb_1),
(J_2,\Jb_2),
(J_3,\Jb_3),
(J_4,\Jb_4)
\right)
=
(\ns+1)(\nsb+1)\times
$$
\beq
\left(
1+{Q_M\over 4}\Bigl[ \alpha'_-
(J_1+J_2+J_3+J_4+1)
+\alpha'_+(\Jb_1+\Jb_2+\Jb_3+\Jb_4+1)\Bigr]
\right)
\label{f4p2}
\eeq
which differs from \ref{f4p1} by the simple factor $(\ns+1)(\nsb+1)$.

Substracting from this total four point function
the first three reducible diagrams of \ref{f4ps},
which are given by sums of the type \ref{calcfeyn},
one gets the one particle irreducible correlator
$$
\epsfbox{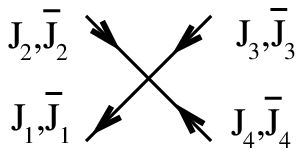} =
\>^{1PI}\!
A^{(4)}_{(\ns,\nsb)}
\left(
(J_1,\Jb_1),
(J_2,\Jb_2),
(J_3,\Jb_3),
(J_4,\Jb_4)
\right)
$$
\beq
=
(\ns+1)(\nsb+1)
{1\over 4}
\left(
-(2+s)
+{Q_M\over 2}(\alpha'_-\ns
+\alpha'_+\nsb)
\right)
\label{f4pirr}
\eeq
which does not depend on the momenta.

It works similarly for the higher order correlators,
with a very quickly growing complexity however.
First of all, every diagram gets more complicated:
instead of the double sum \ref{calcfeyn} the $N$ point diagrams
may include $2(N-3)$ nested sums.
Then, the number of diagrams go from 4 diagrams with  2 different skeletons 
for the four point function (cf \ref{f4ps})
to 26 diagrams of 5 different (oriented) skeletons\footnote{
We say here that two Feynman diagrams
have the same oriented skeleton if they can be ``superposed''
so that the arrows coincide.
So, the diagrams with the same oriented skeleton are obtained
by exchanging the $N-1$ incoming legs ($J_2,..J_N$).
The exchange with the outgoing leg $J_1$ gives
a different oriented skeleton and leads to different computations
because of the different selection rules and directions of internal arrows
(first and second diagrams of \ref{f5ptop} for instance).}
for the five point function,
that we represent here:\footnote{We only draw the five oriented skeletons.}
\beq
\epsfbox{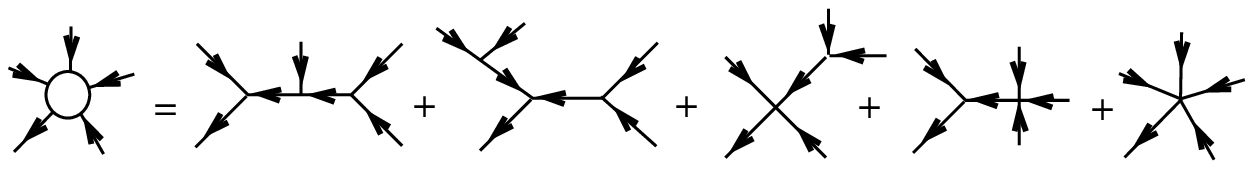},
\label{f5ptop}
\eeq
to 236 diagrams with  12 different oriented skeletons at six points,
2752 diagrams with  34 skeletons at seven points...
The five point function seems already almost out of reach of human
computationnal capabilities.
This is why we used a computer and the ``Mathematica'' program
for symbolic calculation.
We defined an oriented diagram as a tree of which root is the outgoing
momentum.
We generated all of them recursively,
keeping only one diagram with  each different skeleton
in order to speed up computation.
Each of them was then computed recursively,
the result being given by multiple sums.
They could be computed thanks to appropriate rules
for simplification and computation of sums.
And we eventually obtained all the diagrams by symmetrising each
diagram with  a given  skeleton.
We were able to do that up to the six point functions,
which already required some hours of CPU time.
The one particule irreducible five and six point correlators
were then obtained by difference between the total correlators
and these reducible diagrams.
We got for the five point function:
$$
\>^{1PI}\!
A^{(5)}_{(\ns,\nsb)}
\left(
(J_1,\Jb_1),
(J_2,\Jb_2),
(J_3,\Jb_3),
(J_4,\Jb_4),
(J_5,\Jb_5)
\right)
$$
\beq
=
\left(
\,^{\ns+2}_{2}
\right)
\left(
\,^{\nsb+2}_{2}
\right)
\left(
B^{(5)}(\ns,\nsb)
-
\sum_{i=1}^5
(P_i
)^2
\right)
\label{f5pirr}
\eeq
where $B^{(5)}(\ns,\nsb)$ is the following polynomial
of degree 2 in $\ns,\nsb$
$$
B^{(5)}(\ns,\nsb)=
2 + 2\cb+ 2\cn + \cb^2  + 2\cb\cn + cn^2
$$
\beq
 - {{5\,{\it \cn}\,{\it \ns}}\over 3}-{{5\,{\it \cb}\,{\it \nsb}}\over 3}
-{{11\,{\cn^2}\,{\it \ns}}\over {12}}
-{{11\,{\cb^2}\,{\it \nsb}}\over {12}}
+{{{\cn^2}\,{{{\it \ns}}^2}}\over 4}+
   {{{\cb^2}\,{{{\it \nsb}}^2}}\over 4}
+ {{10\,{\it \cb}\,{\it \cn}\,{\it \ns}\,{\it \nsb}}\over 9}
\eeq
$$
\hbox{with }
a_\pm ={Q_M \alpha'_\pm / 4},
$$
and for the six point function
$$
\>^{1PI}\!
A^{(6)}_{(\ns,\nsb)}
\left(
(J_1,\Jb_1),
(J_2,\Jb_2),
(J_3,\Jb_3),
(J_4,\Jb_4),
(J_5,\Jb_5),
(J_6,\Jb_6)
\right)
=
$$
\beq
\left(
\,^{\ns+3}_{3}
\right)
\left(
\,^{\nsb+3}_{3}
\right)
\left[
B^{(6)}(\ns,\nsb)
+
{3\over 2}
\left(
(2+s)-{\cn \ns}-{\cb \nsb}
\right)
\left(
\sum_{i=1}^6
(
P_i )^2
\right)
\right],
\eeq
where $2+s$ comes from $4+2\cn+2\cb$,
with
$$
B^{(6)}(\ns,\nsb)=
{1\over 16}
\left(
-96 - 144\,{\it \cb} - 140\,{\cb^2} - 46\,{\cb^3} -
   144\,{\it \cn} - 280\,{\it \cb}\,{\it \cn} - 138\,{\cb^2}\,{\it \cn}
\right.
$$
$$
-140\,{\cn^2}
 - 138\,{\it \cb}\,{\cn^2} - 46\,{\cn^3} +
   104\,{\it \cn}\,{\it \ns} + 18\,{\it \cb}\,{\it \cn}\,{\it \ns}
+   {\cb^2}\,{\it \cn}\,{\it \ns}
+ 114\,{\cn^2}\,{\it \ns}
$$
$$
  +
   42\,{\it \cb}\,{\cn^2}\,{\it \ns} + 49\,{\cn^3}\,{\it \ns} -
   28\,{\cn^2}\,{{{\it \ns}}^2}
+
   6\,{\it \cb}\,{\cn^2}\,{{{\it \ns}}^2} -
   16\,{\cn^3}\,{{{\it \ns}}^2} + 2\,{\cn^3}\,{{{\it \ns}}^3}
+ 104\,{\it \cb}\,{\it \nsb}
   + 114\,{\cb^2}\,{\it \nsb}
$$
$$
    +
   49\,{\cb^3}\,{\it \nsb} + 18\,{\it \cb}\,{\it \cn}\,{\it \nsb} +
   42\,{\cb^2}\,{\it \cn}\,{\it \nsb} +
   {\it \cb}\,{\cn^2}\,{\it \nsb} -
   104\,{\it \cb}\,{\it \cn}\,{\it \ns}\,{\it \nsb} -
   36\,{\cb^2}\,{\it \cn}\,{\it \ns}\,{\it \nsb} -
   36\,{\it \cb}\,{\cn^2}\,{\it \ns}\,{\it \nsb}
$$
\beq
\left.
 +
   17\,{\it \cb}\,{\cn^2}\,{{{\it \ns}}^2}\,{\it \nsb} -
   28\,{\cb^2}\,{{{\it \nsb}}^2} -
   16\,{\cb^3}\,{{{\it \nsb}}^2} +
   6\,{\cb^2}\,{\it \cn}\,{{{\it \nsb}}^2} +
   17\,{\cb^2}\,{\it \cn}\,{\it \ns}\,{{{\it \nsb}}^2} +
   2\,{\cb^3}\,{{{\it \nsb}}^3}
\right).
\eeq

The polynomials $B^{(5)}$ and $B^{(6)}$ are rather complicated
and by now we have no interpretation of them.
On the other hand, the way these irreducible correlation functions
depend on the momenta is particularly interesting.
First, they are symmetric in all of their legs,
including the outgoing one, which is not the case of the reducible diagram.
Second, they are symmetric under the exchange of $P_i$ into $-P_i$
(or equivalently $J_i,\Jb_i$ into $-J_i-1,-\Jb_i-1$),
as we only have even powers of the momenta (0 or 2).
This is necessary to have a really symmetric correlator
as one leg is outgoing (this was not true for the total correlators,
see \ref{f4p2} e.g.).
This feature was already noticed in the weak coupling regime\cite{DFK},
where higher order correlators where computed at $c=1$ in the case without
screening charges (for matter).
It was obtained there that the irreducible $N$ point functions
only depend on momenta at even power $2$ Int$((N-3)/2)$,
which is similar to the present results in the strong coupling regime
in the case with screenings.
This symmetry of the irreducible correlation functions seems
to be a good check of consistency of this description
by an effective action.
\subsection{Weak coupling regime}

We draw a parallel with the case of weak coupling
in order to show that many things in the previous
derivation are not specific to the strong coupling regime.
Similarly, we use
two copies of the construction of the weak
coupling regime,
one for matter and the other for gravity,
and get the dressed operators  already
introduced in Eqs.\ref{vertexdef},\ref{vcdef}.
One knows from the so-called Seiberg bound that this second
type of dressing is to be chosen only for negative momenta of matter, and this is 
why we reverse the spins in the ``conj'' operators.
This was discussed in details in ref.\cite{DFK},
and it was shown there that the integral representations
give non vanishing results only for one negative momentum,
the others being positive.
We shall simply follow this guide here and not go through the whole discussion
as our main purpose
is simply to show that many aspects
of the previous derivations were not specific of the
strong coupling regime.
So we only consider (at least in a first step) correlators of the type
\beq
\left<
\vertexcw -J_1-1,-\Jhat_1-1,
\vertexw J_2,\Jhat_2,
...
\vertexw J_N,\Jhat_N,
\right>_{\mu_c}.
\eeq
Their scaling law was computed in Eq.\ref{1.5} and
may be rewritten
\beq
\left<
\vertexcw -J_1-1,-\Jhat_1-1,
\vertexw J_2,\Jhat_2,
...
\vertexw J_N,\Jhat_N,
\right>_{\mu_c}
\sim
\mu_c^{\sum_i
\pw_i
-\left({N\over 2}-1\right)
\left(
1+
{\pi\over h}
\right)
}
\label{mudepw}
\eeq
with
\beq
\pw_i
=
-
\left(
J_i+{1\over 2}
\right)
+\left(
\Jhat_i +{1\over 2}
\right)
{\pi\over h}
=
{
\beta_i
\over
\alpha_-
}
+
{1\over 2}
\left(
1+
{\pi\over h}
\right)
=
{
\beta_i
\over
\alpha_-
}
+
{Q\over 2\alpha_-}
\label{Qi}.
\eeq
The three point functions are obtained from previous computations and are
equal to one, up to (different) leg factors:
$$
\left<
\vertexcw -J_1-1,-\Jhat_1-1,
\vertexw J_2,\Jhat_2,
\vertexw J_3,\Jhat_3,
\right>_{\mu_c}
\!\!\!\!\!\!
=
\!\!\!
\epsfbox{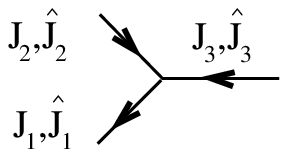}
$$
\beq
= \lfcw J_1, \Jhat_1,
\lfw J_2, \Jhat_2,
\lfw J_3, \Jhat_3,
\
\mu_c^{\pw_1+\pw_2 + \pw_3
-{1\over 2}\left( 1+{\pi\over h}\right)}
\!\!\!
\label{f3p2bw}
\eeq
with
\beq
\lfw J, \Jhat,
=
{1\over
F(1+2J+\pish \Jhat)
}, \qquad 
\lfcw J, \Jhat,
=
{1\over
F(1+2\Jhat+\hspi J)
}
.
\eeq
{}From the particular three point function with
one momentum put to zero,
one deduces the two point function by integration.
The propagator is given by its inverse\footnote{Up to a factor
1/2 here again} and is again given by the moment:
\beq
P_{\mu_c}(J_i,\Jhat_i)=
\pw_i
=
\epsfbox{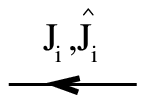}
\label{propw}
\eeq
where from now on we omit the leg factors and the $\mu_c$ dependence.

All these derivations seem really similar to the ones performed
in the case of strong coupling.
And it is actually possible to show that, up to some correpondance,
all the diagrammatic computations are equivalent.
Although the underlying physics is completely different
the following correspondence
\beq
\begin{array}{ccc}
\vertex J_i,\Jb_i, &\to& \vertexw J_i,\Jhat_i, \\
\vertex J_i,\Jb_i, &\to& \vertexw J_i,\Jhat_i, \\
J_i &\to& J_i\\
\Jb_i &\to& \Jhat_i\\
P_i &\to& \pw_i \\
\cn\equiv Q_M\alpha'_-/4 &\to& -1 \\
\cb\equiv Q_M\alpha'_+/4 &\to& \pi/h=\alpha_-^2/2\equiv\rho
\end{array}
\label{corresp}
\eeq
relates both diagramatics.
This can easily be checked for the dependence in $\mu_c$
(Eqs.\ref{mudep3} and \ref{mudepw} respectively\footnote{
Using $(2+s)/2=2-(2-s)/2=2+\cn+\cb\ \to\ 
2+(-1)+(\pi/h)=1+\pi/h$.}),
for the momenta (Eqs.\ref{Pi} and \ref{Qi}),
for the three point function
(Eqs.\ref{f3p2b} and \ref{f3p2bw} respectively),
and for the propagators (Eqs.\ref{prop} and \ref{propw}).
Consequently the higher order correlators and diagrams
obtained either by derivation or from these Feynman rules
obey the same correspondence.
So we get immediately the one particule irreducible correlators
of the weak coupling regime by replacing $\Jb_i,\cn,\cb$
by $\Jhat_i,-1,\rho$ repectively.
The four point one particule irreducible amplitude is now given by
$$
\epsfbox{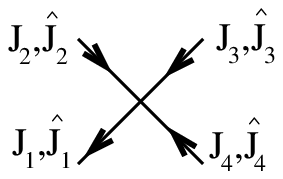} =
\>^{1PI}\!
A^{(4)}_{(\ns,\nshat)}
\left(
(J_1,\Jhat_1),
(J_2,\Jhat_2),
(J_3,\Jhat_3),
(J_4,\Jhat_4)
\right)
$$
\beq
=
(\ns+1)(\nshat+1)
{1\over 2}
\left(
-1-\rho
-\ns
+\rho\nshat
\right)
\label{f4pirrw}
\eeq
where it was used that the first term in \ref{f4pirr} comes from
$-(2+s)/4=-1+(2-s)/4=-1-(\cn+\cb)/2$ which gives via the correspondence
$-1-((-1)+(\pi/h))/2=-(1+\pi/h)/2=-(1+\rho)/2$.
The five point function is
$$
\>^{1PI}\!
A^{(5)}_{(\ns,\nshat)}
\left(
(J_1,\Jhat_1),
(J_2,\Jhat_2),
(J_3,\Jhat_3),
(J_4,\Jhat_4),
(J_5,\Jhat_5)
\right)
$$
\beq
=
\left(
\,^{\ns+2}_{2}
\right)
\left(
\,^{\nshat+2}_{2}
\right)
\left(
B^{(5)}(\ns,\nshat)
-
\sum_{i=1}^5
(\pw_i
)^2
\right)
\label{f5pirrw}
\eeq
with now
\beq
B^{(5)}(\ns,\nsb)=
1 + {{{\it \rho}}^2} + {{3\,{\it \ns}}\over 4} + {{{{{\it \ns}}^2}}\over 4} - 
   {{5\,{\it \rho}\,{\it \nshat}}\over 3} - 
   {{11\,{{{\it \rho}}^2}\,{\it \nshat}}\over {12}} - 
   {{10\,{\it \rho}\,{\it \ns}\,{\it \nshat}}\over 9} + 
   {{{{{\it \rho}}^2}\,{{{\it \nshat}}^2}}\over 4}
\eeq
and for the six point function
$$
\>^{1PI}\!
A^{(6)}_{(\ns,\nshat)}
\left(
(J_1,\Jhat_1),
(J_2,\Jhat_2),
(J_3,\Jhat_3),
(J_4,\Jhat_4),
(J_5,\Jhat_5),
(J_6,\Jhat_6)
\right)
=
$$
\beq
\left(
\,^{\ns+3}_{3}
\right)
\left(
\,^{\nshat+3}_{3}
\right)
\left[
B^{(6)}(\ns,\nshat)
+
{3\over 2}
\left(
2+2\rho+\ns-\rho \nshat
\right)
\left(
\sum_{i=1}^6
(
\pw_i )^2
\right)
\right],
\eeq
with
$$
B^{(6)}(\ns,\nsb)=
{1\over 16}
\left(
  -46 - 2\,{\it \rho} - 2\,{{{\it \rho}}^2} - 46\,{{{\it \rho}}^3} - 39\,{\it \ns} + 
   24\,{\it \rho}\,{\it \ns} - {{{\it \rho}}^2}\,{\it \ns} - 12\,{{{\it \ns}}^2} + 
   6\,{\it \rho}\,{{{\it \ns}}^2}
\right.
$$
$$
- 2\,{{{\it \ns}}^3} +
   87\,{\it \rho}\,{\it \nshat} + 72\,{{{\it \rho}}^2}\,{\it \nshat} + 
   49\,{{{\it \rho}}^3}\,{\it \nshat} + 68\,{\it \rho}\,{\it \ns}\,{\it \nshat} + 
   36\,{{{\it \rho}}^2}\,{\it \ns}\,{\it \nshat} + 
$$
\beq
\left.
   17\,{\it \rho}\,{{{\it \ns}}^2}\,{\it \nshat} - 
   34\,{{{\it \rho}}^2}\,{{{\it \nshat}}^2} - 
   16\,{{{\it \rho}}^3}\,{{{\it \nshat}}^2} - 
   17\,{{{\it \rho}}^2}\,{\it \ns}\,{{{\it \nshat}}^2} + 
   2\,{{{\it \rho}}^3}\,{{{\it \nshat}}^3}
\right).
\eeq

The case without screenings in the case $c=1$ (thus $\rho=1$)
was computed in ref.\cite{DFK} (Appendix B).
The restriction of the previous formulae to this case yields
\beq
\>^{1PI}\!
A^{(4)}_{(\ns,\nshat)}
\left(
(J_1,\Jhat_1),
(J_2,\Jhat_2),
(J_3,\Jhat_3),
(J_4,\Jhat_4)
\right)
=
-1,
\eeq
\beq
\>^{1PI}\!
A^{(5)}_{(\ns,\nshat)}
\left(
(J_1,\Jhat_1),
(J_2,\Jhat_2),
(J_3,\Jhat_3),
(J_4,\Jhat_4),
(J_5,\Jhat_5)
\right)
=
2-\sum_{i=1}^5
(\pw_i
)^2
\eeq
with
$
\pw_i=\Jhat_i-J_1$
in this case,
and
\beq
\>^{1PI}\!
A^{(6)}_{(\ns,\nshat)}
\left(
(J_1,\Jhat_1),
(J_2,\Jhat_2),
(J_3,\Jhat_3),
(J_4,\Jhat_4),
(J_5,\Jhat_5),
(J_6,\Jhat_6)
\right)
=
-6+6\sum_{i=1}^6
(\pw_i
)^2.
\eeq
The correspondence with the results of ref.\cite{DFK} (Eqs.B.1, B.2)
is immediate as the momenta $k_i$ they use are related to ours
by $k_i=\sqrt{2} \pw_i$ in the case $c=1$.
\section{Outlook}
The general conclusion of the present study is that the strong coupling 
physics is governed by  the new cosmological term, very much the way the 
weak coupling arises from the standard Liouville theory. 
Of course a better understanding of the strong coupling physics just described is 
needed. In particular one would like to reach a geometrical understanding of our 
new cosmological term and of its associated area element. A basic point is the deconfinement 
of chirality whose world sheet meaning is still mysterious. One way to gain 
insight would be to study our topological models, and their descriptions of the 
DDK type which should include a pair of chiral bosons of opposite chiralities  
instead of the Liouville field. Our Feynman perturbation involves novel features 
as compared with the previous discussion of ref.\cite{DFK}, but we showed that 
they also appear in the weak coupling regime, once we let the screening numbers fluctuate, 
contrary to ref.\cite{DFK}. We may expect progress in the future.

\end{document}